\documentclass[article,amsmath,11pt,amssymb,floatfix,superscriptaddress,
showpagenumber,nofootinbib]{revtex4}
\usepackage{graphicx}
\usepackage{epsfig}
\usepackage{bm}
\usepackage{amsfonts}

\input epsf

\begin{document}

\title{\Large Modified Holographic energy density driven inflation and some cosmological outcomes}
\author{Gargee Chakraborty}
\affiliation{Department of Mathematics, Amity University, Major
Arterial Road, Action Area II, Rajarhat, New Town, Kolkata 700135,
India.}
\author{Surajit Chattopadhyay}
\email{schattopadhyay1@kol.amity.edu; surajitchatto@outlook.com}
\affiliation{ Department of Mathematics, Amity University, Major
Arterial Road, Action Area II, Rajarhat, New Town, Kolkata 700135,
India.}

\date{\today}

\newpage

\begin{abstract}
\textbf{Abstract}: Motivated by the work of Nojiri et al. \cite{motivation}, the present study reports a model of inflation under the consideration that the inflationary regime is originated by a type of holographic energy density. The infrared cutoff has been selected based on the modified holographic model that is a particular case of Nojiri-Odintsov holographic dark energy \cite{odi1} that unifies phantom inflation with the acceleration of the universe on late-time. On getting an analytical solution for Hubble parameter we considered the presence of bulk viscosity and the effective equation of state parameter appeared to be consistent with inflationary scenario with some constraints. It has also being observed that in the inflationary scenario the contribution of bulk viscosity is not of much significance and its influence is increasing with the evolution of the universe. Inflationary observables have been computed for the model and the slow-roll parameters have been computed. Finally, it has been observed that the trajectories in $n_s - r$  are compatible with the observational bound found by Planck. It has been concluded that the tensor to scalar ratio for this model can explain the primordial fluctuation in the early universe as well.
\textbf{AMSC:} 83F05, 83C99
\end{abstract}

\maketitle
\section{Introduction}
The holographic principle, put forward by \cite{hoof}, has its root in the theory of quantum gravity. It states that the entropy of a system scales with the area of the enveloping horizon, rather than the volume. Inspired by black hole thermodynamics, the holographic principle gives a connection of the short cutoff of a quantum field theory to a long distance cutoff due to the limit set by the formation of a black hole \cite{oliver}. This consideration has been applied in the study of the late time acceleration of the universe to a considerable extent. In a cosmological framework of the late time universe, the vacuum energy constitutes a dark energy (DE) sector having a holographic origin. This DE is called holographic dark energy (HDE) \cite{li}. During recent years, a considerable amount of efforts have been devoted to the development of the various cosmological aspects and generalizations of HDE. Literatures in this arena include \cite{hde1,hde2,hde3,hde4}. A detailed review on HDE has been presented in \cite{li}. Applying holographic principle to HDE in a universe with a characteristic length scale $L$ and reduced Planck mass $M_p$ the following expression can be derived (see \cite{li}):
\begin{equation}\label{1}
\rho_{de}=C_1 M_p^4+C_2 M_p^2 L^{-2}+C_3 L^{-4}+...
\end{equation}
The first term being disfavoured due to its incompatibility with holographic principle and the vacuum fluctuation estimated from UV-cutoff quantum field theory being $\rho_{de}\sim \Lambda^4\lesssim M_p^2 L^{-2}$, it has been prescribed that the above expression should begin from the second term. The third and subsequent terms being negligible in comparison with the second term, the expression for HDE has come out to be \cite{li}
\begin{equation}\label{hde}
\rho_{de}=3C^2 M_p^2 L^{-2}
\end{equation}
Here, $C$ is an arbitrary parameter. For the remaining part of the paper we will consider $M_p^2=1$. In a recent paper, Chattopadhyay et al. \cite{chatto} reported an inflationary universe in $f(T)$ framework through slow-roll formalism and holographic Ricci dark energy as its driving force.

In the present paper we will probe the holographic inflation. The inflation theory proposes a period of extremely rapid (exponential) expansion of the universe during its first few moments. It is developed by Alan Guth, Andrei Linde, Paul Steinhardt and Andy Albrecht, offers solutions to these problems and several other open questions in cosmology. It states that prior to the more gradual Big Bang expansion, there was a period of extremely rapid (exponential) expansion of the universe, during which the time energy density of the universe was dominated by a cosmologist constant type of vacuum energy that later decayed to produce the matter and radiation that fill the universe today. Inflation was both rapid and strong. Inflation is now considered an extension of the Big Bang theory since it explains the above puzzle so well while retaining the basic assumption of a homogeneous expanding universe \cite{hde4}. Rest of the paper is organized as follows: In Section II we have demonstrated the inflation with modified holographic density as its driving force. In this section, we have derived an analytical solution for the inflationary scale factor. In a subsection under section II, we have discussed the role of bulk viscosity in the modified holographic inflation. In another subsection we have discussed the slow roll parameters from the Hubble parameter derived under the inflationary settings through modified holographic density. Lastly in this section we have considered some limiting scenarios. We have concluded in Section III.

\section{Holographic Inflation}
In this section, we will construct the basic model of holographic inflation. We consider a homogeneous and isotropic Friedmann-Robertson-Walker (FRW) geometry with metric
\begin{equation}\label{frw}
dS^2 = - dt^2 + a(t)\left(\frac{dr^2}{1-k^2 r^2} + r^2 d{\Omega}^2\right)
\end{equation}
where, $a(t)$ is scale factor and $k=0,+1,-1$ correspond to flat, close and open universe. In the present work, we consider a flat universe i.e. $k=0$.

As already stated, the primary focus of the present work is to demonstrate the inflation driven by an effective fluid of holographic origin. We denote the effective fluid density responsible for driving the inflation as $\rho_{inf}$.

As the inflation driving fluid is of holographic origin, we consider its source as modified holographic dark energy (MHDE), where the IR cutoff is a linear combination of $H^2$ and $\dot{H}$ \cite{b2,b3}

\begin{equation}\label{a}
\rho_{\Lambda} = \frac{2}{\alpha - \beta}\left[\dot{H} + \frac{3 \alpha}{2} H^2\right]
\end{equation}
It may be noted that  holographic Nojiri-Odintsov DE proposed in \cite{odi1} is the most general HDE model and the holographic inflation follows naturally from that proposal. The form (\ref{a}) is a specific example of the HDE proposed in \cite{odi1}. In studying the inflationary scenario, we will take $\rho_{\Lambda}$ as $\rho_{inf}$ where $\rho_{inf}$ is the energy density of the effective field that derives the inflation and having the possibility of originating from a scalar field, modified theory of gravity or from any other sources \cite{H1, H2, H3, H4}. Here, ${L^{-2}_{IR}} = \dot{H} + \frac{3 \alpha}{2} H^2$. As we are considering an inflationary scenario, we incorporate a correction to IR cutoff because of quantam effect $L = \sqrt{{L^2_{IR}} + \frac{1}{{\Lambda^2_{UV}}}}$. This approach has already been attempted in \cite{motivation}

Friedmann's first equation in inflationary scenario is
\begin{equation}\label{b}
H^2 = \frac{1}{3} \rho_{inf}
\end{equation}
Following Elizalde and Timoshkin \cite{eli}, the conservation equation can be written as
\begin{equation}\label{b}
\dot{\rho}_{inf}+3H(\rho_{inf}+p_{eff})=0
\end{equation}
where, $p_{eff}=p_{inf}+\Pi$, where $\Pi$ represents the bulk viscous pressure. As we are considering MHDE as the driving force for the inflation, we have $\rho_{\Lambda}=\rho_{inf}$ in Eq. (\ref{b}). Hence, using Eqns. (\ref{a}) and (\ref{b}), we get the solution for reconstructed Hubble parameter $H$ as
\begin{equation}\label{aa}
H=\sqrt{a^{-3 \alpha +\frac{3 (\alpha -\beta )}{\Lambda _{{UV}}^4}} {C_1}+\frac{2 \Lambda _{{UV}}^2}{3 \left(-\alpha
+\beta +\alpha  \Lambda _{{UV}}^4\right)}}
\end{equation}
As in Eq.(\ref{b}) we have $H$ expressed in terms of $a$, we can express the $\rho_{inf}$ as
\begin{equation}\label{aaa}
\rho_{inf} = 3 \left({C_1} \left(2^{-1/m} \left(\frac{-{C_1}+e^{2 {C_2} f m+\sqrt{f} m t}}{f}\right)^{\frac{1}{m}}\right)^{-3
\alpha +\frac{3 (\alpha -\beta )}{\Lambda _{{UV}}^4}}+\frac{2 \Lambda _{{UV}}^2}{3 \left(-\alpha +\beta +\alpha  \Lambda _{{UV}}^4\right)}\right)
\end{equation}

Using $H=\frac{\dot{a}}{a}$ we have the following analytical solution for reconstructed scale factor $a(t)$ as
\begin{equation}\label{c}
a=2^{-1/m} \left(\frac{-{C_1}+e^{\sqrt{f} m t+2 f m {C_2}}}{f}\right)^{\frac{1}{m}}
\end{equation}
where $m = 3 \alpha -\frac{3 (\alpha -\beta )}{\Lambda _{{UV}}^4}$ and $f = \frac{2 \Lambda _{{UV}}^2}{3 \left(-\alpha +\beta +\alpha  \Lambda _{{UV}}^4\right)}$. Since $\Lambda _{{UV}}$ is neither equal to $0$ nor to infinity, Eq. (\ref{c}) provides the evolution of scale factor. If $C_1$ is very small, then it does not contribute significantly to the expression $-{C_1}+e^{\sqrt{f} m t+2 f m {C_2}}$ and hence we can write $\left(-{C_1}+e^{\sqrt{f} m t+2 f m {C_2}}\right)^{\frac{1}{m}}\approx e^{\left(\sqrt{f}  t+2 f  {C_2}\right)}$. Hence, from (\ref{c}) we can write $a(t)\approx a_0 e^{H_0 t}$, which is the de Sitter solution for scale factor with $a_0=\left(\frac{1}{2f}\right)^{1/m}e^{2fC_2}$ and $H_0=m\sqrt{f}$. Hence, this consideration is consistent with the basic inflationary features.

\subsubsection{Consideration of bulk viscous pressure}
Ability of bulk viscosity to drive inflationary expansion is being discussed since eighties. References \cite{bulk1,bulk2} argued in favour of the bulk viscosity to drive inflation. However, \cite{bulk3} argued that it is the non perturbative effect that is responsible for the negative pressure during the inflationary expansion and ruled out bulk viscosity as a potential candidate for being the driving fore behind the inflationary expansion of the early universe. In a later study, ref.\cite{bulk4} argued for a non linear generalization of causal linear thermodynamics that could describe viscous inflation without particle production. In a very recent work, Bamba and Odintsov \cite{bulk5} demonstrated a fluid model having EoS that includes bulk viscosity and argued that a fluid description of inflation has an equivalence with the description of inflation in terms of scalar field theories. They \cite{bulk5} also demonstrated the realization of graceful exit from inflation for the reconstructed models of fluid.

Inspired by the work of \cite{bulk5} we incorporate the bulk viscous pressure as time varying form to view the consequences if the inflationary fluid includes bulk viscosity. As a specific case the bulk-viscous pressure is considered as $\Pi=-3H\xi$, where $\xi=\xi_0+\xi_1 \frac{\dot{a}}{a}+\xi_2 \frac{\ddot{a}}{a}=\xi_0 + \xi_1 H + \xi_2 (\dot{H}+ H^2)$, where $\xi_0$, $\xi_1$ and $\xi_2$ are positive constants. Under the consideration of $\rho_{\Lambda}=\rho_{inf}$ we get reconstructed $\xi$ and $\Pi$ as
\begin{equation}\label{d}
\xi =\xi_0 + {\xi_1} \sqrt{a^{-3 \alpha +\frac{3 (\alpha -\beta )}{\Lambda _{UV}^4}} {C_1}+\frac{2 \Lambda
_{UV}^2}{3 \left(-\alpha +\beta +\alpha  \Lambda _{{UV}}^4\right)}}+ \xi_2 \left(a^{-3 \alpha +\frac{3 (\alpha -\beta )}{\Lambda
_{{UV}}^4}} {C_1}+\frac{2 \Lambda _{{UV}}^2}{3 \left(-\alpha +\beta +\alpha  \Lambda _{{UV}}^4\right)}\right)
\end{equation}
and
\begin{equation}\label{e}
\begin{array}{c}
\Pi =-3 \sqrt{a^{-3 \alpha +\frac{3 (\alpha -\beta )}{\Lambda _{{UV}}^4}} {C_1}+\frac{2 \Lambda _{
{UV}}^2}{3 \left(-\alpha
+\beta +\alpha  \Lambda _{{UV}}^4\right)}} \\  \left({\xi_0}+{\xi_1} \sqrt{a^{-3 \alpha +\frac{3 (\alpha -\beta )}{\Lambda _{{UV}}^4}}
{C_1}+\frac{2 \Lambda _{{UV}}^2}{3 \left(-\alpha +\beta +\alpha  \Lambda _{{UV}}^4\right)}} + {\xi_2} \left(a^{-3 \alpha +\frac{3
(\alpha -\beta )}{\Lambda _{{UV}}^4}} {C_1}+\frac{2 \Lambda _{{UV}}^2}{3 \left(-\alpha +\beta +\alpha  \Lambda _{{UV}}^4\right)}\right)\right)
\end{array}
\end{equation}
The absolute of reconstructed bulk viscous pressure in Eq.(\ref{e}) is plotted in Fig.\ref{visp} against redshift $z$  for various combinations of $\alpha$ and $\beta$ with $\alpha-\beta>0$ as well as $<0$.

In inflationary scenario, the conservation equation in presence of bulk-viscous pressure $\Pi$  is
\begin{equation} \label{f}
\dot{\rho}_{inf} + 3 H (\rho_{inf} + p_{inf} + \Pi) =0
\end{equation}
In the Eq.(\ref{f}), putting the value of $\rho_{inf}$ from Eq.(\ref{aaa}), $\Pi$ from Eq.(\ref{e}), $H$ from Eq.(\ref{aa}), we get thermodynamic pressure as $p_{inf}$

\begin{equation}\label{g}
\begin{array}{c}
p_{inf} = \frac{3 a^{-3 \alpha +\frac{3 (\alpha -\beta )}{\Lambda _{{UV}}^4}} {C_1} \left(-\alpha +\beta +(-1+\alpha ) \Lambda
_{{UV}}^4\right)}{\Lambda _{{UV}}^4}-\frac{2 \Lambda _{{UV}}^2}{ \beta -\alpha +\alpha  \Lambda _{{UV}}^4}+ \\ 3 \sqrt{a^{-3 \alpha
+\frac{3 (\alpha -\beta )}{\Lambda _{{UV}}^4}} {C_1}+\frac{2 \Lambda _{{UV}}^2}{3 \left( \beta -\alpha +\alpha  \Lambda _{{UV}}^4\right)}}
\left({\xi_0}+a^{-3 \alpha +\frac{3 (\alpha -\beta )}{\Lambda _{{UV}}^4}} {C_1} {\xi_2}+\frac{2 {\xi_2} \Lambda _{{UV}}^2}{3
\left( \beta -\alpha +\alpha  \Lambda _{{UV}}^4\right)}+ \right. \\ \left. {\xi_1} \sqrt{a^{-3 \alpha +\frac{3 (\alpha -\beta )}{\Lambda _{{UV}}^4}}
{C_1}+\frac{2 \Lambda _{{UV}}^2}{3 \left( \beta -\alpha +\alpha  \Lambda _{{UV}}^4\right)}}\right)
\end{array}
\end{equation}

Equation of state parameter (EoS) is $w_{eff}= \frac{p_{inf} + \Pi}{\rho_{inf}}$. Hence in this equation putting the value of $p_{inf}$ from Eq.(\ref{g}), $\Pi$ from Eq.(\ref{e}) and the value of $\rho_{inf}$ from Eq.(\ref{aaa}), we get $w_{eff}$ as

\begin{equation}\label{h}
w_{eff} = \frac{\frac{3 {C_1} \left(2^{-1/m} \left(\frac{-{C_1}+e^{2 {C_2} f m+\sqrt{f} m t}}{f}\right)^{\frac{1}{m}}\right)^{-3 \alpha
+\frac{3 (\alpha -\beta )}{\Lambda _{{UV}}^4}} \left( \beta -\alpha +(-1+\alpha ) \Lambda _{{UV}}^4\right)}{\Lambda _{{UV}}^4}-\frac{2
\Lambda _{{UV}}^2}{ \beta -\alpha +\alpha  \Lambda _{{UV}}^4}}{3 \left({C_1} \left(2^{-1/m} \left(\frac{-{C_1}+e^{2 {C_2} f
m+\sqrt{f} m t}}{f}\right)^{\frac{1}{m}}\right)^{-3 \alpha +\frac{3 (\alpha -\beta )}{\Lambda _{{UV}}^4}}+\frac{2 \Lambda _{{UV}}^2}{3
\left( \beta -\alpha +\alpha  \Lambda _{{UV}}^4\right)}\right)}
\end{equation}

Eq. (\ref{h}) gives us that $w=-1.00817,~-1.00816,~-1.00812,~-1.00807$ for $t=0,~0.01,~0.05,~0.1$ respectively. Thus, $w_{eff}=w_{inf}\approx -1$. Hence, the introduction of bulk viscosity having its consequence consistent with inflation \cite{lote}. Furthermore, Fig.\ref{visp} makes it apparent that in the very early phase of the universe $\Pi$ has contribution close to $0$. However, with exit from inflation the contribution of bulk viscous pressure is increasing with the evolution of the universe.

\subsubsection{Slow roll parameters}
As we have obtained an analytic solution for $H$, it is now possible to obtain the Hubble slow roll parameter $\epsilon_n$, where $n$ is positive integer.  The Hubble slow roll parameter is defined as \cite{motivation}
\begin{equation}
\epsilon_{n+1} = \frac{d ln |\epsilon_n|}{dN}
\end{equation}
where with $\epsilon_0 \equiv H_{ini}/H$ and $N \equiv ln(a/a_{ini})$ the e-folding number, and
where $a_{ini}$ and $H_{ini}$ is the scale factor at the beginning of inflation and
the corresponding Hubble parameter (inflation ends when $\epsilon_1 =1$).
Therefore, we can find  the values of the inflationary observables,
namely the scalar spectral index of the curvature perturbations $n_s$ ,
its running $\alpha_s $, the tensor spectral index $n_T$ and the tensor-to-scalar ratio
$r$ \cite{ma1}.
The slow-roll parameters in terms of Hubble parameter are \cite{motivation}
\begin{equation}\label{gg}
\epsilon_1 = - \frac{\dot{H}}{H^2}
\end{equation}
\begin{equation}\label{ggg}
\epsilon_2 = \frac{\ddot{H}}{H \dot{H}} - \frac{2 \dot{H}}{H^2}
\end{equation}
\begin{equation}\label{gggg}
\epsilon_3 = (\ddot{H} H - 2 {\dot{H}}^2)^{-1}\left[\left(\frac{H\dot{H}\dddot{H}-\ddot{H}({\dot{H}}^2 + H \ddot{H})}{H \dot{H}}\right)-\frac{2\dot{H}}{H^2}(H \ddot{H}-2 {\dot{H}}^2)\right]
\end{equation}

\begin{figure}
\begin{minipage}{14pc}
\includegraphics[width=16pc]{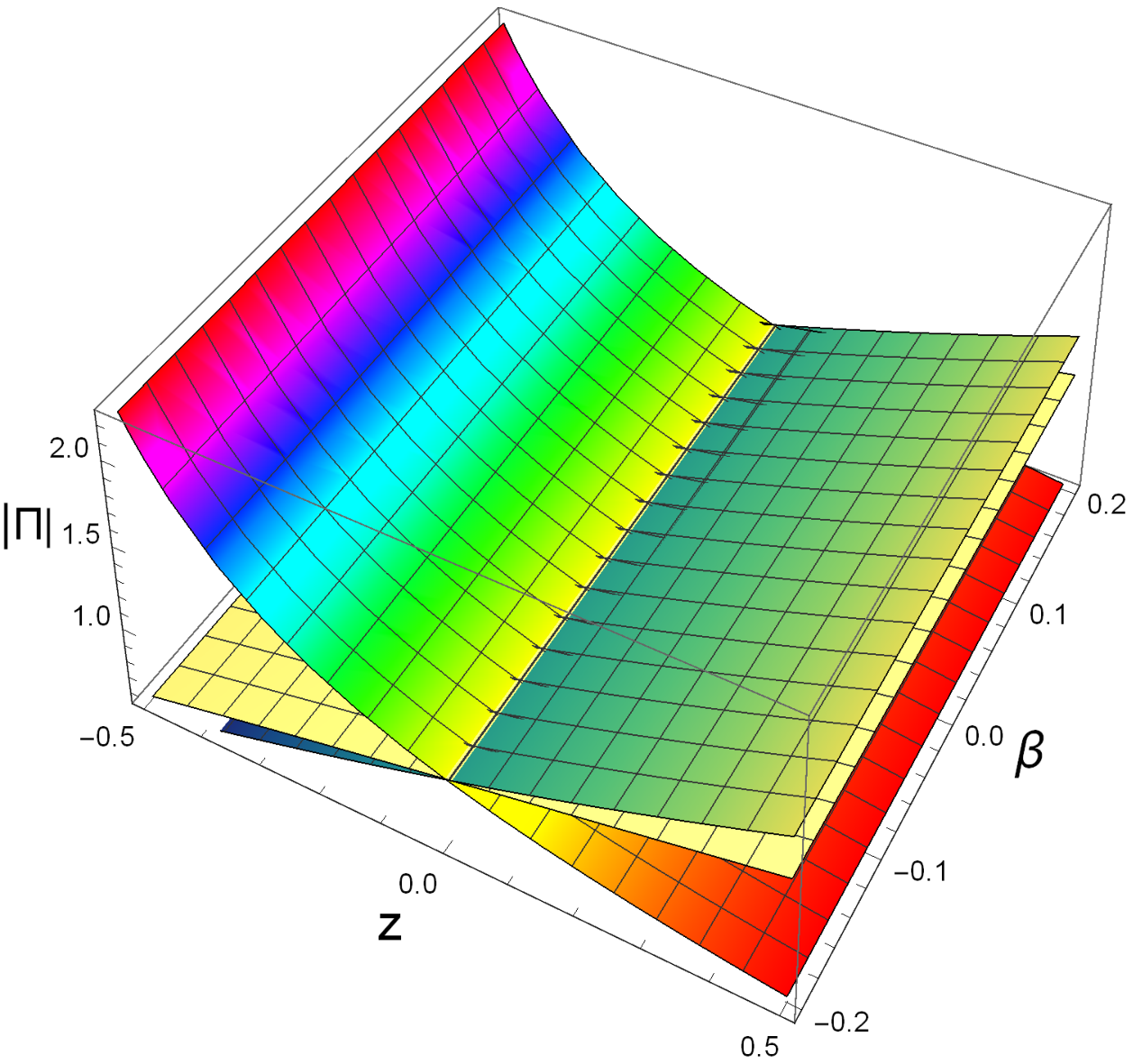}
\caption{Evolution of reconstructed Bulk Viscous pressure (see Eq. (\ref{e})) against red-shift $z$. We have taken $\xi_0 = 0.52$, $\xi_1=0.33$, $\xi_2=0.22$, $C_1=0.24$, $\Lambda_{UV}=26$, $\beta \in [-0.2,+0.2]$ with $\alpha=-0.6,~0.6$ and $0.4$.}
\label{visp}
\end{minipage}\hspace{3pc}%
\begin{minipage}{14pc}
\includegraphics[width=16pc]{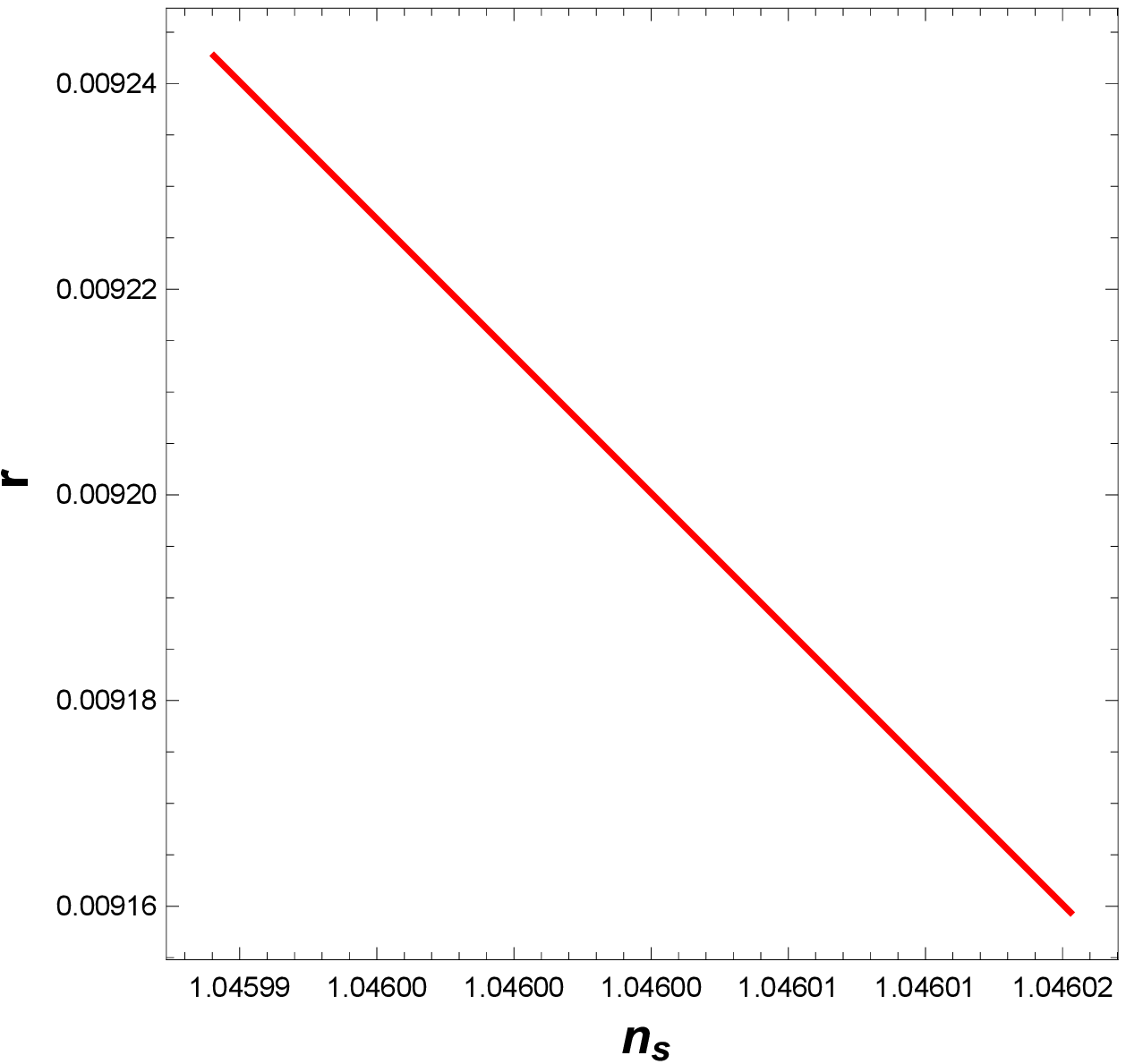}
\caption{Evolution of tensor to scalar ratio $r$ against spectral index $n_s$. We have taken $\Lambda_{UV}=24,~C_1=0.024,~C_2=0.48$. Lines have been drawn for $\{\alpha=0.0077,\beta=180\}$, $\{\alpha=0.008,\beta=190\}$ and $\{\alpha=0.0084,\beta=200\}$. The lines are almost coincident.}
\label{str}
\end{minipage}\hspace{3pc}%
\end{figure}

Using Eq.(\ref{aa})in Eqns. (\ref{gg}), (\ref{ggg})and (\ref{gggg}), the slow roll parameters comes out to be:
\begin{equation}\label{i}
\epsilon_1 = \frac{9 {C_1} \left(-\alpha +\beta +\alpha  \Lambda _{{UV}}^4\right)^2}{2 \Lambda _{{UV}}^4 \left(3 {C_1} (-\alpha
+\beta )+2 \left(2^{-1/m} \left(\frac{-{C_1}+e^{2 {C_2} f m+\sqrt{f} m t}}{f}\right)^{\frac{1}{m}}\right)^{3 \alpha -\frac{3 (\alpha -\beta
)}{\Lambda _{{UV}}^4}} \Lambda _{{UV}}^2+3 {C_1} \alpha  \Lambda _{{UV}}^4\right)}
\end{equation}

\begin{equation}\label{j}
\epsilon_2 = -3 \alpha +\frac{3 (\alpha -\beta )}{\Lambda _{{UV}}^4}+\frac{9 {C_1} \left(-\alpha +\beta +\alpha  \Lambda _{{UV}}^4\right)^2}{\Lambda
_{UV}^4 \left(3 {C_1} (-\alpha +\beta )+2 \left(2^{-1/m} \left(\frac{-{C_1}+e^{2 {C_2} f m+\sqrt{f} m t}}{f}\right)^{\frac{1}{m}}\right)^{3
\alpha -\frac{3 (\alpha -\beta )}{\Lambda _{{UV}}^4}} \Lambda _{{UV}}^2+3 {C_1} \alpha  \Lambda _{{UV}}^4\right)}
\end{equation}

\begin{equation}\label{k}
\begin{array}{c}
\epsilon_3 = \left[{9 {C_1} \left(2^{-1/m} \left(\frac{-{C_1}+e^{2 {C_2} f m+\sqrt{f} m t}}{f}\right)^{\frac{1}{m}}\right)^{\frac{3
\alpha }{\Lambda _{{UV}}^4}} \left(-\alpha +\beta +\alpha  \Lambda _{{UV}}^4\right)^2}\right]\\ \times \left[\Lambda _{{UV}}^4 \left(-3 {C1} \left(2^{-1/m}
\left(\frac{-{C1}+e^{2 {C2} f m+\sqrt{f} m t}}{f}\right)^{\frac{1}{m}}\right)^{\frac{3 \alpha }{\Lambda _{{UV}}^4}} (\alpha -\beta
)+ \right. \right.\\ \left. 2 \left(2^{-1/m} \left(\frac{-{C_1}+e^{2 {C_2} f m+\sqrt{f} m t}}{f}\right)^{\frac{1}{m}}\right)^{3 \left(\alpha +\frac{\beta }{\Lambda
_{{UV}}^4}\right)} \Lambda _{{UV}}^2+3 {C_1} \left(2^{-1/m} \left(\frac{-{C_1}+e^{2 {C_2} f m+\sqrt{f} m t}}{f}\right)^{\frac{1}{m}}\right)^{\frac{3
\alpha }{\Lambda _{{UV}}^4}} { \alpha  \Lambda _{{UV}}^4}\right]^{-1}
\end{array}
\end{equation}
Putting the value of $\epsilon_1$, $\epsilon_2$ respectively from Eq.(\ref{i}),Eq.(\ref{j}) respectively in $n_s = 1-2\epsilon_1-2\epsilon_2$, we get spectral index $n_s$ as
\begin{equation}\label{m}
\begin{array}{c}
n_s = 1-\frac{9 {C_1} \left(-\alpha +\beta +\alpha  \Lambda _{{UV}}^4\right){}^2}{\Lambda _{{UV}}^4 \left(3 {C_1}
(-\alpha +\beta )+2 \left(2^{-1/m} \left(\frac{-{C_1}+e^{2 {C_2} f m+\sqrt{f} m t}}{f}\right)^{\frac{1}{m}}\right)^{3 \alpha -\frac{3 (\alpha
-\beta )}{\Lambda _{{UV}}^4}} \Lambda _{{UV}}^2+3 {C_1} \alpha  \Lambda _{{UV}}^4\right)} \\ -2 \left(-3 \alpha +\frac{3 (\alpha -\beta
)}{\Lambda _{{UV}}^4}+\frac{9 {C_1} \left(-\alpha +\beta +\alpha  \Lambda _{{UV}}^4\right){}^2}{\Lambda _{{UV}}^4 \left(3 {C_1}
(-\alpha +\beta )+2 \left(2^{-1/m} \left(\frac{-{C_1}+e^{2 {C_2} f m+\sqrt{f} m t}}{f}\right)^{\frac{1}{m}}\right)^{3 \alpha -\frac{3 (\alpha
-\beta )}{\Lambda _{{UV}}^4}} \Lambda _{{UV}}^2+3 {C_1} \alpha  \Lambda _{{UV}}^4\right)}\right)
\end{array}
\end{equation}
Putting the value of $\epsilon_1$, $\epsilon_2$, $\epsilon_3$ respectively from Eq.(\ref{i}),Eq.(\ref{j}), Eq.(\ref{k}) respectively in $\alpha_s = -2 \epsilon_1 \epsilon_2 - \epsilon_2 \epsilon_3$, we get $\alpha_s$ as
\begin{equation}
\begin{array}{c}
\alpha_s = \left[{108 {C_1} \left(2^{-1/m} \left(\frac{-{C_1}+e^{2 {C_2} f m+\sqrt{f} m t}}{f}\right)^{\frac{1}{m}}\right)^{3 \left(\alpha
+\frac{\alpha +\beta }{\Lambda _{{UV}}^4}\right)} \left(-\alpha +\beta +\alpha  \Lambda _{{UV}}^4\right)^3}\right]\\ \times \left[{\Lambda _{{UV}}^6 \left(2
\left(2^{-1/m} \left(\frac{-{C_1}+e^{2 {C_2} f m+\sqrt{f} m t}}{f}\right)^{\frac{1}{m}}\right)^{3 \left(\alpha +\frac{\beta }{\Lambda _{{UV}}^4}\right)}
\Lambda _{{UV}}^2+ \right. }\right. \\ \left. \left. 3 {C_1} \left(2^{-1/m} \left(\frac{-{C_1}+e^{2 {C_2} f m+\sqrt{f} m t}}{f}\right)^{\frac{1}{m}}\right)^{\frac{3
\alpha }{\Lambda _{{UV}}^4}} \left(-\alpha +\beta +\alpha  \Lambda _{{UV}}^4\right)\right)^2\right]^{-1}
\end{array}
\end{equation}
Putting the value of $\epsilon_1$ from Eq.(\ref{i}) in $n_T = - 2 \epsilon_1$, we get tensor spectral index $n_T$ as
\begin{equation}\label{nT}
n_T = -\frac{9 {C_1} \left(-\alpha +\beta +\alpha  \Lambda _{{UV}}^4\right)^2}{\Lambda _{{UV}}^4 \left(3 {C_1} (-\alpha
+\beta )+2 \left(2^{-1/m} \left(\frac{-{C_1}+e^{2 {C_2} f m+\sqrt{f} m t}}{f}\right)^{\frac{1}{m}}\right)^{3 \alpha -\frac{3 (\alpha -\beta
)}{\Lambda _{{UV}}^4}} \Lambda _{{UV}}^2+3 {C_1} \alpha  \Lambda _{{UV}}^4\right)}
\end{equation}
Putting the value of $\epsilon_1$ from Eq.(\ref{i})in $r=16\epsilon_1$, we get tensor to scalar ratio $r$ as
\begin{equation}\label{l}
r = \frac{72 {C_1} \left(-\alpha +\beta
+\alpha  \Lambda _{{UV}}^4\right)^2}{\Lambda _{{UV}}^4 \left(3 C_1 (-\alpha +\beta)+2 \left(2^{-1/
m} \left(\frac{-{C_1}+
e^{2 {C_2} f m + \sqrt{f} m t}}{f}\right)^{\frac{1}{m}}\right)^{3 \alpha -\frac{3 (\alpha -\beta )}{\Lambda
_{{UV}}^4}} \Lambda _{{UV}}^2+3 {C_1} \alpha  \Lambda _{{UV}}^4\right)}
\end{equation}
The evolution of tensor to scalar ratio $r$ in Eq.(\ref{l}) is plotted against spectral index $n_s$ in Eq.(\ref{m}) in Fig.\ref{str}. It is observed that the trajectories in $n_s - r$ plane exhibit a decreasing behaviour, which is consistent with the Jawad et al.\cite{jd} observation. Here, $r<0.168$ (95$\%$ CL, Planck TT + LowP),which is the observational bound found by Planck. Hence, our calculated tensor to scalar ratio for this model is consistent with the observational bound of Planck. Hence, it can explain the primordial fluctuation in the early universe.

Now, we will deduce the particle horizon $R_P$ and the event horizon $R_E$. $R_P$ is the maximum distance from which light could have travelled to the observer in the age of the universe and the event horizon $R_E$ is the largest comoving distance from which light emitted now can ever reach the observer in future.It is the boundary beyond which the light cannot affect the observer. An event horizon is an acknowledged feature of the expanding universe \cite{co2,co1}.
The particle horizon $R_P$ and the event horizon $R_E$ is given by the respective differential equations:
\begin{equation}\label{DE1}
\dot{R_P}=H {R_P} + 1
\end{equation} and
\begin{equation}\label{DE2}
\dot{R_E}= H {R_E} - 1
\end{equation}
We know Hubble parameter $H=\frac{\dot{a}}{a}$. Using the expression of $a$ from Eq.(\ref{c}), we get $H$ as $
H=\left(1+\frac{{C_1}}{-{C_1}+e^{2 {C_2} f m+\sqrt{f} m t}}\right) \sqrt{f}$
Now putting this value of $H$ in Eqns. (\ref{DE1}) and (\ref{DE2})and solving the equations we get
\begin{equation}
R_P = \left(-{C_1}+e^{2 {C_2} f m+\sqrt{f} m t}\right)^{\frac{1}{m}} {C_3}-\frac{\left(1-{C_1} e^{-2 {C_2}
f m-\sqrt{f} m t}\right)^{\frac{1}{m}} {2F1}\left[\frac{1}{m},\frac{1}{m},1+\frac{1}{m},{C_1} e^{-2 {C_2} f m-\sqrt{f}m t}\right]}{\sqrt{f}}
\end{equation}
\begin{equation}
R_E=\left(-{C_1}+e^{2 {C_2} f m+\sqrt{f} m t}\right)^{\frac{1}{m}} {C_4}+\frac{\left(1-{C_1} e^{-2 {C_2} f m-\sqrt{f}
m t}\right)^{\frac{1}{m}} {2F1}\left[\frac{1}{m},\frac{1}{m},1+\frac{1}{m},{C_1} e^{-2 {C_2} f m-\sqrt{f} m t}\right]}{\sqrt{f}}
\end{equation}
If $\Lambda_{UV}\rightarrow \infty$ then $f \rightarrow 0$ and $m\rightarrow 3\alpha$. Hence in that case $R_E$ and $R_P$ will tend to infinity.

\subsubsection{Limiting $\Lambda_{UV} \rightarrow \infty$}
Ultraviolet cutoff $\Lambda_{UV}$ has a major role in understanding the inflation. Here we will study the parameters when there is no ultraviolet cutoff i.e., $\Lambda_{UV}\rightarrow\infty$. In this case, scale factor $a$ calculated in Eq.(\ref{c}) will also tend to infinity. As $a=\frac{1}{1+z}$ which implies that $z\rightarrow-1$ which further implies that there is a scope of exit from inflation. In this case, Hubble parameter $H$ in Eq.(\ref{aa}) will take the value $\sqrt{C_1 a^{-3\alpha}}$. As $H$ is real so $C_1\geq0$. In this case $\rho_{inf}$ in Eq.(\ref{aaa}) tends to $0$ and also $p_{inf}$ tends to $0$. Effective EoS $w_{eff}$ will take the value $\alpha-1$ as $\Lambda_{UV}\rightarrow\infty$. As in the inflationary scenario $w\approx -1$, we may infer that $\alpha$ should be very small.

\section{Concluding remarks}
Motivated by the works of Nojiri et al. \cite{motivation} and Oliveros and Acero \cite{oliver} the present study attempted to study a model of inflation under the consideration that the inflationary regime is originated by a type of holographic energy density as in references \cite{b2,b3}. The holographic density is basically a The infrared cutoff has been selected based on the modified holographic model considered here. Because of the high energy scales in the inflationary regime the infrared cutoff has been corrected by ultraviolet cutoff $\Lambda_{UV}$. This procedure helped us in getting an analytical solution for Hubble parameter, which in turn could give us a solution for scale factor (see Eqs. (\ref{aa})) and (\ref{c}). Afterwards we considered the presence of bulk viscosity $\Pi$ and the effective equation of state parameter appeared to be consistent with inflationary scenario with the necessity that $C_1$ and $\alpha$ are small.

The $C_1$, an integration constant and $\alpha$, a constant parameter of the infrared cutoff are already thoroughly explained in the previous sections. As $\alpha$ is required to be small, in Fig. \ref{visp} we have chosen values of $|\alpha|<1$. The $\beta$ has been varied in such a manner that it can get hold of the various combinations of $\{\alpha,\beta\}$ leading to positive and negative differences. In Fig. \ref{visp} it is observed that for $\alpha-\beta<0$ (the surface having upwards growth with $z$) the bulk viscosity is very low in the inflationary scenario and subsequently it s increasing. Hence, for this combination, it is observed that in the inflationary scenario the contribution of bulk viscosity is not of much significance and its influence is increasing with the evolution of the universe. It is consistent with our assumption that in the inflationary scenario only a fluid of holographic origin is considerable and other components are non-existent. The low bulk viscosity is indicative of the absence of other fluids in the inflationary phase. However, we can further generate the conditions on $\alpha$ and $\beta$ to derive the role of bulk viscosity in the early inflationary scenario. If we consider Eq. (\ref{e}), we can have a further insight into $\alpha$ and $\beta$. Hence, in Fig. \ref{visp} we generate two more surfaces that indicate decreasing pattern of bulk viscosity. However, for this case with $\alpha-\beta>0$ the bulk viscosity in the inflationary scenario is not very high in magnitude.
 
Based on the analytical solution of $H$, the inflationary observables have been computed for the present model in Eqs. (\ref{m}) to (\ref{nT}), and the slow-roll parameters have been computed in Eqs. (\ref{gg}) to (\ref{k}). Additionally, it has been observed that the trajectories in $n_s - r$ plane exhibit a decreasing behaviour, which is consistent with the Jawad et al.\cite{jd} observation. Also, it has been found that $r<0.168$ (95$\%$ CL, Planck TT + LowP),which is the observational bound found by Planck. Hence, our calculated tensor to scalar ratio for this model is consistent with the observational bound of Planck. Hence, it can explain the primordial fluctuation in the early universe as well.

While concluding, we would like to state that we have incorporated bulk viscosity in the inflationary scenario driven by holographic energy density. Although it has been observed that in a holographic fluid driven inflation the contribution of bulk viscosity is negligible, it has also been observed that the effect of bulk viscosity is increasing with expansion of the universe. This approach can further be extended to the reconstruction of Starobinsky inflation, and also to modified gravity like $f(T)$ and $f(G)$ framework.

\section{Acknowledgement}
Authors are thankful to the anonymous reviewer for the constructive suggestions. Surajit Chattopadhyay acknowledges financial support from the Council of Scientific and Industrial Research (Government of India) with Grant No. 03(1420)/18/EMR-II. Both the authors acknowledge the facility and hospitality provided by the Inter-University Centre for Astronomy and Astrophysics (IUCAA), Pune, India during a scientific visit from December, 2019 till January, 2020.

\end{document}